\documentclass[twocolumn,prb,aps,showpacs,preprintnumbers,amsmath,amssymb,superscriptaddress,floatfix,footinbib]{revtex4}%twocolumn,,preprint
\usepackage{graphicx}% Include figure files
\usepackage{dcolumn}% Align table columns on decimal point
\usepackage{bm}% bold math
\usepackage{subfigure}
\usepackage{sidecap}
\begin{document}
\title{Continuous corrections to the molecular Kohn-Sham gap
and virtual orbitals}
\author{H\'{e}ctor Mera}
\affiliation{Niels Bohr Institute and Nano-Science Center, Universitetsparken 5,DK-2100 Copenhagen. Denmark}
\date{\today}
\begin{abstract}
We use projector operators 
to correct the Kohn-Sham Hamiltonian of density functional theory (KS-DFT) so that the resulting mean-field
scheme yields, in finite systems, virtual orbitals and energy gaps in better agreement with those predicted by
quasiparticle theory. The proposed correction term is a scissors-like operator of the form 
$(\hat{I}-\hat{\rho})\delta \hat{H}(\hat{I}-\hat{\rho})$, where $\hat{I}$ is the identity operator, $\hat{\rho}$  the density matrix of 
the N-particle system and $\delta \hat{H}$ 
is either the difference between the N+1- and N-particle Kohn-Sham Hamiltonians or a non-self-consistent approximation to it. 
Such a term replaces the Kohn-Sham virtual orbitals of the $N$-particle system by the HOMO and virtual orbitals
of the system with $N+1$ particles in an attempt to mimic a true quasiparticle spectrum. Using a local density approximation (LDA)
we compute the gaps of a variety of small molecules finding good agreement with experiment and computationally more demanding methods. 
For these systems we examine the physical origin of this gap correction and show that so-called band gap discontinuity, $\Delta_{xc}$,
contains electrostatic contributions that do not originate from the discontinuity in the exchange-correlation potential. The similarity between the corrected and Hartree-Fock virtual orbitals is
illustrated and the extent to which the bare LDA virtual orbitals are improved is considered. The lack of band-gap discontinuity 
and the presence of self-interaction errors in the proposed correction are also discussed.
\end{abstract}
\maketitle

\section{Introduction}
The accurate determination of the fundamental energy gap in many-electron systems, such as molecules and solids, 
by means of a computationally inexpensive mean-field 
method remains an open problem despite of recent progress\cite{Cappellini97,Johnson98,Malloci2004,EG,rosselli-2006,KKRPM,Cehovin08}. The calculation of fundamental energy gaps based on the relevant 
eigenvalues of the 
Kohn-Sham Hamiltonian of Density Functional Theory (KS-DFT)
\cite{HK,KS} severely underestimates 
the experimental energy gap both in finite and extended  systems \cite{Perdew82,Sham83,sham,Godby88}. KS-DFT has been extremely successful in the computation of ground state averages for which accurate
density functionals are known, such as the total energy and magnetic moment \cite{parryang,Jones89,rwg62}. This success is also due to 
the extremely good
ratio between the accuracy and the computational demands of this method; KS-DFT not only yields accurate results for these quantities, but it
is also very fast and simple to implement for most available exchange-correlation (XC) functionals. 

Unfortunately KS-DFT is not designed to give
addition and removal energies in a single calculation: the Kohn-Sham eigenvalues have no direct physical interpretation except for the ones
associated with the highest occupied molecular orbital (HOMO) \cite{coabarth}; none of the Kohn-Sham orbitals has a known physical meaning;
the Kohn-Sham virtual orbitals do not enter the self-consistent cycle and play no role in the theory;
and the Kohn-Sham Hamiltonian is unphysical as we will show clearly below. 

In KS-DFT the ground state N-particle electron
density is obtained by the self-consistent solution of the Kohn-Sham equations, which are characterized by a non-interacting Hamiltonian of
the the form
\begin{equation}
H^{\mathrm{KS}}_N=-\frac{\nabla^2}{2}+v_s[n_N](\mathbf{r}) \, ,
\label{eq:HKS}
\end{equation}
where the effective Kohn-Sham potential, $v_s$, is a functional of the electron density of the $N$-particle system, $n_N$, and includes the external and Hartree potentials 
as well as the XC potential, which remains unkown and is notoriously difficult to approximate within the realm of KS-DFT
since  it is a discontinuous functional of the density \cite{Sham83,Perdewgap} and must remove the fictitious self-interaction terms present in the Hartree term. 

Note that $v_s$ in Eq.(\ref{eq:HKS}) is the same for all of the Kohn-Sham orbitals, whether occupied or not. This fact is 
in clear contrast with the Hartree-Fock (HF) equations which in the basis of eigenstates of the 
HF Hamiltonian, $\psi_i$, can be written as
\begin{equation}
h_0(\mathbf{r})\psi_j(\mathbf{r})+\sum_{j}^{\mathrm{occ}}V_{jj}\psi_i(\mathbf{r})-\sum_{j}^{\mathrm{occ}}V_{ij}\psi_j(\mathbf{r})=\epsilon_i\psi_i(\mathbf{r})\,
,
\label{eq:HF}
\end{equation} 
where $h_0(\mathbf{r})$ accounts for the kinetic energy and external potential, $V_{ii}(\mathbf{r})=\int d^3\mathbf{r'} |\mathbf{r}-\mathbf{r'}|^{-1}|\psi_i(\mathbf{r'})|^2$ and 
$V_{ij}(\mathbf{r})=\int d^3\mathbf{r'} \psi_j(\mathbf{r'})|\mathbf{r}-\mathbf{r'}|^{-1}\psi_i(\mathbf{r'})$. Since
$\delta_{i,j}=\langle \psi_i|\psi_j\rangle$, when $\psi_i$ is unoccupied the electrostatic fields that enter the HF equations are
generated by $N$ particles while they are generated by $N-1$ particles when $\psi_i$ is occupied\cite{Cehovin08}. This is precisely what one would expect from simple electrostatic considerations
of a charged particle added to or removed from a dielectric or metallic system. While the HF approximation has many
drawbacks associated with the lack of dynamical correlation, this one physically appealling feature 
is totally absent in KS-DFT. Note that in many-body perturbation theory, the one-particle Green's function in the frequency representation
has poles at the excitation energies of the $N+1$ ($N-1$) electron system for unoccupied (occupied) states \cite{HedinLundqvist,negele,Nozieres66}.

This is one reason, other than the discontinuity in the XC potential, why exact KS-DFT underestimates 
the experimental gap in finite systems, while reproducing exact electron density as well as the total and HOMO energies. 
Consider the concrete 
application to finite and spherically symmetric systems such as atoms. The exact KS-DFT effective potential 
is self-interaction free and decays asymptotically into the vacuum as\cite{KKRPM,Almbladh84} $\propto
-1/r$, where $r$ is the radial distace to the atomic nucleus, \emph{both for unoccupied and occupied states}. However, in 
quasiparticle theory  this potential should decay as\cite{rwg64,Rinkephd} $\propto -1/r^4$  for unoccupied states 
(particle added to the system). The $\propto -1/r^4$ tail results from the interaction between the added electronic charge and the
polarization charge density it induces in the spherical symmetric system. Since $-1/r^4$ is more confining than $-1/r$, the KS-DFT LUMO energy 
underestimates the experimental one\footnote{as obtained from the system's electron affinity}, contributing to the underestimation of the gap. 
These arguments are purely electrostatic in nature, involving only continuous functionals of the density that
vanish with vanishing fractional particle numbers \emph{or in infinite systems}. In other words the effects discussed above 
are proportional to the added charge and vanish with zero 
added charge and thus are not directly related to the well-known discontinuity in the XC potential. To summarize:
the effective potential of KS-DFT makes no distinction between occupied and virtual orbitals and, as a result, the screened electrostatic 
repulsion of the added electron by all the other electrons is missing in KS-DFT. 

This band-gap problem also arises in KS-DFT-based calculations of electronic transport through molecular conductors\cite{Cehovin08,Ke07}.
In Ref.~\onlinecite{Ke07} it is shown that exact-exchange DFT predicts conductances that are one order of magnitude smaller than those
obtained by the Hartree-Fock approximation, despite of the fact that the electronic densities obtained by both methods are the same. An
additional important problem --not addressed here-- in the context of quantum electronic transport is the presence of self-interaction errors 
that arise through the use of approximated XC functionals\cite{Ke07,Toher05}. In Refs.~\onlinecite{Cehovin08,Ke07,Toher05} the importance
of the single-particle orbitals is also highlighted. Atomistic simulations of electronic transport are computationally demanding and a
single-particle theory which is computationally inexpensive, free of self-interaction errors, able to accurately predict the molecular
HOMO-LUMO gap and --somehow-- able to provide \emph{good} single-particle orbitals is needed.

The rest of this paper is organized as follows: in the next section we discuss the correction proposed by Cehovin et al.\cite{Cehovin08},
which should correct the above-mentioned deficiency of KS-DFT; in Section III we present our numerical results for the HOMO-LUMO gaps and
virtual orbitals of a few selected molecules. We end with the conclusions in Section IV. 

\section{Correction to the Kohn-Sham Hamiltonian} 
These arguments show that to improve the gap values, the $N$-particle Kohn-Sham
Hamiltonian needs to be modified in order to include information about singly-charged state with $N + 1$ electrons. 
We note that the so-called $\Delta\mathrm{SCF}$ method, where
the gap is calculated by means of three total energy calculations as $E_g=E_{N+1}+E_{N-1}-2E_N$, is known to provide
better gaps than those obtained directly from the Kohn-Sham eigenvalues. Since, for all $N$, the electron affinity (EA) of the $N$-particle system 
and the ionization potential (IP) of the $N+1$-particle system satisfy:
\begin{equation}
EA(N)=IP(N+1)\,,
\label{eq:eaip}
\end{equation}
the fundamental gap can be obtained  by means of two calculations of the Kohn-Sham HOMO, one for the N-particle system and 
another for the (N+1)-particle system \cite{Cappellini97,Malloci2004,EG,rosselli-2006}. 

To account for these differences between the Kohn-Sham and quasiparticle Hamiltonians, here we apply a variant of the method of 
\emph{improved virtual orbitals} \cite{collins,collinsPRA,goddard}. 
Given the density matrix of the $N$-electron system, $\hat{\rho}_N$, we define a projector over the unoccupied states of 
the $N$-electron system, $\hat{I}-\hat{\rho}_N$  and a mean-field-like Hamiltonian
\begin{equation}
\hat{H}_N\equiv\hat{H}^{KS}_N+(\hat{I}-\hat{\rho}_N)\delta \hat{H}(\hat{I}-\hat{\rho}_N)
\label{eq:KS-sym}
\end{equation}
where $\delta \hat{H}$ is defined below. Because of the action of $\hat{I}-\hat{\rho}_N$, $\delta H$ acts only on the Kohn-Sham 
virtual orbitals of the $N$-particle system, \emph{leaving the ground state properties invariant}. The occupied orbitals of
Eq.~(\ref{eq:KS-sym}) are just the occupied Kohn-Sham orbitals. The virtual orbitals depend on how $\delta \hat{H}$ is defined.
In applications of the so-called \emph{scissors operator} method \cite{scissors},  $\delta
\hat{H}=\epsilon\hat{I}$, where $\epsilon$ is typically a free parameter. Following
Refs.~\onlinecite{Cappellini97,Malloci2004,EG,rosselli-2006}, its can be defined as
$\epsilon \equiv \epsilon_H^{\mathrm{N+1}}-\epsilon_L^{\mathrm{N}}$ , and its action is just to shift the virtual-orbital energies 
of the $N$-electron system 
so that Eq.(\ref{eq:eaip}) is satisfied. Here $\epsilon_{H(L)}^{\mathrm{N+1(N)}}$ is the HOMO (LUMO) energy 
of the Kohn-Sham N+1 (N) particle  system.  Such an approach corrects the virtual orbital energies but leaves the orbitals themselves unchanged. 
Hence to obtain a  mean-field-like approximation to the unoccupied quasiparticle orbitals, we define $\delta \hat{H}$ as
\begin{equation}
\delta \hat{H}_N=\hat{H}^{KS}_{N+1}-\hat{H}^{KS}_{N} \, .
\label{eq:correction}
\end{equation}
The above equation takes into account the fact that the electromagnetic fields acting on accupied and unoccupied orbitals are 
generated by different numbers of particles. Eqs.(\ref{eq:KS-sym}-\ref{eq:correction}) replaces the virtual 
orbitals of the $N$-particle system by an approximation to the HOMO and virtual orbitals of the $N+1$ particle system, hence accounting for 
repulsion of the added particle by all the other particles, as well as some approximated form of screening given by the approximation used 
for XC potential in the Kohn-Sham Hamiltonians in Eq.(\ref{eq:correction}).  
\section{Numerical Results}
\subsection{Computational Details}
We calculate the HOMO-LUMO gap and LUMO's of several isolated molecules, by means
of Eqs.(\ref{eq:KS-sym}-\ref{eq:correction}). In our calculations we approximate the XC potential 
by means of a spin-unpolarized  LDA \footnote{Other local and semi-local functionals were used, yielding quantitatively similar results and
identical qualitative trends.}, where the correlation part is given as in Ref.~\onlinecite{vosko}. We use either the self-consistent (SCF) LDA density 
matrix of the $N+1$ particle system or a non-self-consistent approximation (NSCF) to it,  
 $\hat{\rho}_{N+1} \approx \hat{\rho}_N+|\psi^{N}_L \rangle\langle \psi^{N}_L |$,
where $\psi^N_L$ is the LUMO of the $N$-particle system. We use an open source quantum chemistry package \cite{pyquante} which 
we have tested by comparing the results of our
calculations (HOMO energies and HOMO-LUMO gaps) with those reported in Ref.~\onlinecite{NIST} as well as 
independent calculations using output from the GAMESS code\cite{gamess}. From these experiences we estimate our numerical 
error bar in the orbital energies to be $\sim$ 1-2 mHa for LDA-based calculations and typically less than 1 mHa for HF calculations. 
We use the gaussian basis set 6-31G** and perform our calculations at the 
GGA geometries given in Ref.~\onlinecite{NIST}. The geometry of $Na_4$ was taken from Ref.~\onlinecite{na4geom}.
\subsection{Gaps}
Calculated HOMO-LUMO gaps using different methods are presented in Table \ref{table:gaps}, which clearly demonstrates that both 
the SCF and NSCF methods greatly improve the value of the gap, which is much closer to both the HF and $\Delta$SCF gaps 
relative to the bare LDA gaps, in line with previous calculations that use essentially the same method 
\cite{Cappellini97,Malloci2004,EG,rosselli-2006}. In most cases  both the SCF and NSCF methods yield gaps that are somewhat 
smaller than the HF ones. 
In addition the SCF is shown to produce gaps that are smaller than those of the NSCF. It is worth pointing out that the NSCF method 
is essentially as computationally expensive as the LDA, requiring only one extra iteration for the (N+1)-particle system and some extra 
post-processing, yet the gaps are greatly improved.
\begin{table}[t]%[!ht]
{
  \centering
  \begin{tabular}{|c|d|d|d|d|d|}
    \hline\hline
  Molecule&\multicolumn{1}{c|}{LDA}&\multicolumn{1}{c|}{HF}&\multicolumn{1}{c|}{$\Delta$SCF}&\multicolumn{1}{c|}{SCF}&\multicolumn{1}{c|}{NSCF}\\
    \hline\hline
    $\mathrm{H_2}$&12.2&22.5&23.6&20.6&21.2\\
    \hline
    $\mathrm{Li_2}$&1.4&5.2&5.6&4.9&5.2\\
    \hline
    $\mathrm{LiH}$&2.9&8.3&9.3&6.8&7.3\\
    \hline
    $\mathrm{H_2O}$&7.3&19.1&18.0&14.8&15.6\\
    \hline
     $\mathrm{CH_4}$&11.6&21.6&20.2&17.6&18.2\\
    \hline
    $\mathrm{Na_2}$&1.2&4.6&5.1&4.4&4.9\\
    \hline
   $\mathrm{Na_4}$&0.4&3.7&4.2&3.4&3.9\\
    \hline\hline
  \end{tabular}
  \par}
\caption{\label{table:gaps} Calculated HOMO-LUMO gaps in electro-volts for several small molecules. Both
the SCF and the NSCF approximations produce gaps that are smaller and much closer to the HF and $\Delta$SCF values than the LDA ones. The
experimental values for $\mathrm{LiH}$, $\mathrm{Na_2}$ and $\mathrm{Na_4}$ are 7.6, 4.5 and 3.4 $eV$, respectively. }
\end{table}

\begin{figure}
\begin{center}
\includegraphics[scale=0.6]{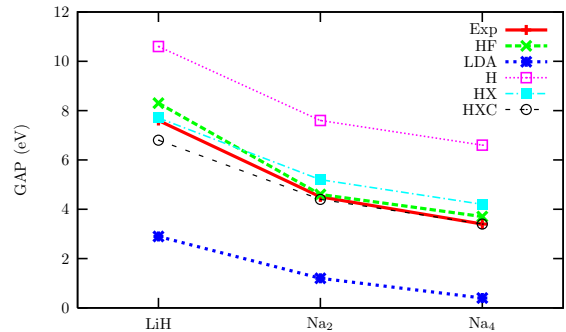}
\caption{(color online) For the molecules shown, we compare the Hartree (H),Hartree-exchange (HX) 
contributions to the SCF gap
correction (HXC) with the experimental (Exp), HF and LDA gaps (see text). All the lines are guides
to the eye. The Hartree-only correction overestimates the experimental gap.
Adding both exchange and correlation to the correction lowers the gap value,
bringing it closer to the HF and experimental values obtained from Ref.~\onlinecite{webbook}. 
\label{fig:hxc}}
\end{center}
\end{figure}

In Fig.\ref{fig:hxc} we consider separately the Hartree, and LDA exchange and correlation contributions to the SCF gap correction for 
three of these molecules. The Hartree contribution is obtained from the difference between the Hartree potentials, $v_h$, 
of the $N+1$ and $N$ particle system, by means of  Eqs.(\ref{eq:KS-sym}-\ref{eq:correction}) with 
$\delta \hat{H}=\hat{v}_{h}[\hat{\rho}_{N+1}]-\hat{v}_{h}[\hat{\rho}_{N}]$, and similarly for the exchange and
correlation LDA potentials. For the systems considered the main effect is associated to the purely electrostatic correction
which increases the LDA gap by a large ammount ($\sim$ 7-8  eV for the molecules shown), overestimating both the Hartee-Fock 
and experimental gaps. Both the exchange and correlation contributions to the correction given by
Eqs.(\ref{eq:KS-sym}-\ref{eq:correction}) lower the value of the gap (by about 3-4 eV for the molecules shown), 
improve the agreement with the experiment and are consistent with a simple screening picture.
As argued above, these corrections are not included in the N-particle Kohn-Sham gap and hence are included 
in the so-called band-gap discontinuity, $\Delta_{xc}$, which 
is typically defined as the difference between the exact and Kohn-Sham gaps. In finite systems 
$\Delta_{xc}$, should not be confused with the discontinuity in the XC potential, since, in  addition to the discontuinity, 
there are continuous contributions originating for both the Hartree and XC potentials. In other words, the
term \emph{band-gap discontinuity} is not appropriate for finite systems. The discontinuity in the
XC potential is not included in our calculations yet the effects we describe with Eq.(\ref{eq:correction}) are not included
in KS-DFT.
\subsection{Virtual Orbitals}
An interesting question is whether the application of Eqs.(\ref{eq:KS-sym}-\ref{eq:correction}) yields virtual orbitals in better agreement with quasiparticle
orbitals.  There is no precise way to evaluate the quality of a mean field orbital, but since the adoption
of Eqs.(\ref{eq:KS-sym}-\ref{eq:correction}) was based  originally based
on the property of the HF Hamiltonian discussed in the introduction, we take the HF LUMO as a reference state and compute its quantum-mechanical overlap with the LDA, SCF and NSCF
LUMO's. In addition to this overlap we evaluate the following distance
\begin{equation}
\sqrt{\sum_{i} (|c^{HF}_{L,i}|-|c_{L,i}|)^2}
\label{eq:dist}
\end{equation}
where $i$ labels a basis orbital, $c^{HF}_{L,i}$ is the i-th expansion coefficient of the HF LUMO, and $c_{L,i}$ is the corresponding
coeffient of  LDA, SCF and NSCF LUMO's. Eq.\ref{eq:dist} gives another measure of the similarity between the HF and LDA, SCF and NSCF LUMO
expansion coefficients.
The results of this comparison  are shown in table \ref{table:overlaps}. 
We note that the excellent agreement 
between the HF and the bare LDA LUMO's is improved in the molecules shown, except for the NSCF calculation of the sodium tetramer, 
where the HF LUMO overlaps with the NSCF LUMO+1 instead of the the LUMO. 

\begin{table}[ht]
{
  \centering
  \begin{tabular}{|c|c|c|c|c|c|c|}
    \hline\hline
   &\multicolumn{3}{|c|}{Overlap}&\multicolumn{3}{|c|}{Eq.(\ref{eq:dist})}\\
   \cline{2-7}
    \raisebox{1.5ex}{Molecule} &\multicolumn{1}{c|}{LDA}&\multicolumn{1}{c|}{SCF}&\multicolumn{1}{c|}{NSCF}&\multicolumn{1}{c|}{LDA}&\multicolumn{1}{c|}{SCF}&\multicolumn{1}{c|}{NSCF}\\
    \hline\hline
    $\mathrm{H_2}$&0.998&1.000&1.000&0.050&0.003&0.012\\
    \hline
    $\mathrm{Li_2}$&0.975&0.999&1.000&0.144&0.050&0.011\\
    \hline
    $\mathrm{LiH}$&0.972&0.997&0.997&0.122&0.024&0.026\\
    \hline
    $\mathrm{H_2O}$&0.994&0.999&1.000&0.057&0.019&0.007\\
    \hline
    $\mathrm{CH_4}$&0.997&0.999&0.996&0.050&0.016&0.006\\
    \hline
    $\mathrm{Na_2}$&0.955&0.996&0.999&0.174&0.056&0.021\\
    \hline
    $\mathrm{Na_4}$&0.948&0.985& 0.000&0.114&0.058&0.131\\
    \hline\hline
  \end{tabular}
  \par}
\caption{\label{table:overlaps} Second, third and fourth columns: absolute values of the overlaps between the Hartree-Fock LUMO and the LUMO's obtained by means of the 
LDA, SCF, and NSCF.  Fifth, sixth, and seventh columns: distance (as given by Eq.(\ref{eq:dist})) between the HF and LDA, SCF, and NSCF
LUMO states, respectively. }
\end{table}
While promising we do not expect these results to be general, due to the  approximated nature of our calculations. 
The XC functionals considered are continuous functionals of the density which are not self-interaction free. This has
consequences both for the validity of our approach in infinite systems and for the values of the LUMO energy obtained for finite systems. 
Since the functionals we consider are continuous, one would expect the correction represented by Eq.(\ref{eq:correction}) to vanish 
for infinite systems. For the small molecules considered the electrostatic effect is large and its influence in the virtual orbitals
seems to be captured by means of Eqs.(\ref{eq:KS-sym}-\ref{eq:correction}), unfortunately this will  not be the case for both 
extended systems or large molecules. The present approach does not modify the occupied orbitals, 
which would be identical to those of underlying KS-DFT method, i.e., the calculated HOMO energies are too high because of the self-interaction
error present in the functionals considered. This leads to LUMO energies that are also too high when compared with experiments. 
In finite systems, the LUMO energy could in principle be improved by considering functionals that are self-interaction free 
like, e.g., those of proposed in Refs.~\onlinecite{Toher05,filispal}.
\section{Conclusions}
In conclusion, we have proposed a simple correction to the virtual orbitals of the KS-DFT Hamiltonian 
that approximates the screened electrostatic repulsion of the added particle by the other electrons.
For the molecules tested, the method systematically improves on the calculated fundamental gaps while provinding also improved LUMOs. Due
to the various approximations used we do not regard the observed LUMO improvement as general. 
The non-self-consistent version of the proposed correction improves
over the KS-DFT LDA without a significant increase in computational time and difficulty of implementation. The proposed mean-field might be
useful to improve the accuracy of KS-DFT-based calculations of the resistance of molecular junctions, since these have proven to be highly
sensitive both to the gap and the details of the mean field orbitals\cite{Ke07, Cehovin08}.  Alternatively the proposed correction might be
used to produce a better starting point for calculations of molecular systems based on many-body perturbation theory.

\emph{Acknowledgements} HM ackowledges support from the Danish Reasearch Agency, NABIIT project ``Materials design using grid technology''.
We gratefully acknowledge useful discussions with the members of Rex Godby's group in York. We are thankful to Jan Jensen, Kurt Stokbro and 
Thomas Bondo for many useful discussions and suggestions.

\end{document}